\documentstyle[12pt,epsfig]{article}

\newcommand{\NP}[1]{ Nucl.\ Phys.\ {#1}}

\newcommand{\PR}[1]{Phys.\ Rev.\ { #1}}
\newcommand{\PRL}[1]{ Phys.\ Rev.\ Lett.\ { #1}}

\newcommand{\be}{\begin{equation}}
\newcommand{\ee}{\end{equation}}
\newcommand{\ba}{\begin{eqnarray}}
\newcommand{\ea}{\end{eqnarray}}

\newcommand{\da}{^\dagger}

\newcommand{\f}{\frac}

\newcommand{\dd}{\displaystyle}

\newcommand{\lr}[1]{{ \left( \, #1 \, \right) }}
\newcommand{\gs}{g'_5 \, \! ^2}
\newcommand{\ima}{{\mbox{Im}\,}}

\begin{document}
\baselineskip=20pt

\begin{center}
  {\Large  \bf Strong tree level unitarity violations 
in the extra dimensional Standard Model 
with scalars in the bulk} 

\vspace{.5cm} { Stefania DE CURTIS$^a$,
Daniele DOMINICI$^{a}$, Jos\'e R. PELAEZ$^{a,b}$
}

\emph{ $^a$ INFN,
Sezione di Firenze and
Dip. di Fisica, Univ. degli Studi, Firenze,
Italy.}

\emph{ $^b$Dept. de F\'{\i}sica Te\'orica II,
  Univ. Complutense, 28040 Madrid, Spain.}
\end{center}

\vspace{-.5cm}
\noindent
\rule{\textwidth}{.1mm}
\begin{abstract}
\noindent
We show how the tree level unitarity  violations 
of compactified extra dimensional extensions of the Standard Model
become much stronger when the scalar sector is included in the bulk.
This effect occurs when the couplings are not
suppressed for larger Kaluza-Klein levels, and could have
relevant consequences for the phenomenology of the next generation 
of colliders.
We also introduce a simple and generic formalism
to obtain unitarity bounds for finite energies, taking
into account coupled channels  including the towers 
of Kaluza-Klein excitations.
\end{abstract}
\vspace{-.5cm}
\rule{\textwidth}{.1mm}

\section{Introduction}

The presence of extra compact dimensions is a common 
feature of the unification of gravity with strong and electroweak
interactions.
Recent theories 
include  realizations where  the Standard Model (SM) 
interactions feel some of these compact extra  dimensions whose 
scales are in the TeV
range \cite{Antoniadis:1990ew}. 
After 
dimensional reduction to four dimensions these models contain 
towers of Kaluza-Klein
(KK) excitations of the gluon, of the $W$, $Z$, of the photon
and possibly of the Higgs and of the fermions with masses in the TeV
range.
This makes these models testable
at present and planned accelerators.
Lower bounds from the electroweak
precision data on the compactification scale of these models, when
fermions are localized on the brane or in different points of the
bulk, are in the range of 2-5 TeV
\cite{Masip:1999mk}. These bounds become much weaker when
all particles live in the bulk, known as universal extra dimensions models
\cite{Barbieri:2000vh,Appelquist:2000nn}. 

In general in higher dimensional theories one expects a violation of tree
level unitarity at high energies. Therefore an explicit estimate
of the unitarity bounds of the theory is important to understand the
validity of tree level calculations.
Recently the tree-level unitarity of   unbroken 
five dimensional (5D)
Yang Mills theories has been  shown to hold at low energy
\cite{SekharChivukula:2001hz}, proving some cancellations  among
contributions of KK excitations of different  levels.
Furthermore a
theorem  analogous 
to the standard Equivalence Theorem (ET) \cite{ET,Lee:1977eg},
that relates at high energies the longitudinal components of gauge bosons to 
their associated Goldstone bosons (GB), was also demonstrated. In the unbroken
extra dimensional Yang Mills case, what has been shown is the
equivalence of 
longitudinal KK gauge bosons $V^\mu_{L\,(n)}$ and their
corresponding $V^5_{(n)}$ components of the 5D
gauge fields.
The equivalence theorem has been shown to hold also in the  case
of spontaneously broken 5D extensions of the SM \cite{DeCurtis:2002nd}. 
In such a case, the ET has allowed 
to calculate the 
$W^{+}_{L\,(m)}W^{-}_{L\,(n)}\rightarrow W^{+}_{L\,(p)}W^{-}_{L\,(q)}$ 
scattering amplitudes and show that the partial wave unitarity limit on
 the mass of the Higgs receives only very tiny corrections from the pure 
Kaluza Klein gauge sector. 

In this work we will see, however, that 
the tree level unitarity bounds on the Higgs mass, can be drastically modified
due to the interactions with the KK modes of the scalar sector
when the scalar interactions are not sufficiently suppressed for each
KK level. In section 2, we revisit the unitarity constraints
in the coupled channel partial wave formalism, providing
a simple generalization of the most familiar unitarity bound.
Next we introduce in section 3 the 5D SM with a Higgs field in the bulk
that will be used in section 4
to illustrate the use of this bound in a simple case.
In section 5 we discuss the effect of adding more extra dimensions.

\section{Unitarity bounds with coupled channels}

Let us show how to obtain unitarity bounds 
from scattering amplitudes when many different two particle 
states are accessible. As it is well known \cite{scattering}, the $S$ 
matrix unitarity relation  $S S^\dagger=1$
translates into simple relations for the elements of the $T$
matrix $T_{\alpha\beta}$, where $\alpha,\beta,...$ denote 
the different states physically available. These relations are even 
simpler if the matrix elements are projected into partial waves.
For two-scalar states, 
which are the ones that we will be considering here, they
are defined as
\be
t_{\alpha\beta}^{J}(s)=\frac{1}{32\pi}\int^1_{-1}\,d(\cos\theta) 
T_{\alpha\beta}(s,t,u)
P_J(\cos\theta),
\ee
where $\theta$ is the angle between the first and third three-momenta and
 $P_J$ is  the $Jth$ Legendre polynomial.
Let us recall that any given two body state $\alpha$
 should carry a $1/\sqrt{2}$
normalization factor if its two particles are identical.
We are using scalar fields to show how to obtain bounds 
from coupled channel unitarity, but the generalization
to particles with spin is straightforward \cite{scattering}, once the
partial waves have been defined in terms of the total angular momentum,
combining their spins and the spatial angular momentum.
Another reason to choose scalar fields is that
tree level unitarity violation occurs more rapidly in the
Higgs-Goldstone boson sector than in other sectors of the SM. 

In particular, with these definitions, and
if there is only one two-body accessible state, $\alpha$,  each partial
wave $t^J_{\alpha\alpha}$ satisfies the following simple unitarity relation 
\begin{equation}
\ima t^J_{\alpha\alpha}  = \sigma_\alpha \,\vert
 \,t^J_{\alpha\alpha} \vert ^2 \quad,
\label{uni1}
\end{equation}
where $\sigma_\alpha$ is the phase space available
for the state $\alpha$, given by $\sigma_\alpha=2 q_\alpha/\sqrt{s}$ 
and $q_\alpha$ is the C.M. momentum of
the state $\alpha$. 
As a matter of fact, this
relation only holds from the $\alpha$ threshold up to the energy where
the next state, $\beta$, is physically accessible.  Above that point,
if $\beta$ is another two-particle state,
the unitarity relation for the partial waves can be written
as
\begin{equation}
\left.
\begin{array}[h]{l}
\ima t^J_{\alpha\alpha} =\sigma_\alpha \,\vert\, 
t^J_{\alpha\alpha} \vert ^2+
\sigma_\beta \,\vert\, t^J_{\alpha\beta} \vert ^2\\ \\
\ima t^J_{\alpha\beta} = \sigma_\alpha \, 
t^J_{\alpha\alpha}\,  (t^J_{\alpha\beta})^*+
\sigma_\beta \,t^J_{\alpha\beta}\, (t^J_{\beta\beta})^*\\ \\
\ima t^J_{\beta\beta} = \sigma_\alpha \,\vert\, t^J_{\alpha\beta} \vert ^2+
\sigma_\beta \,\vert\, t^J_{\beta\beta} \vert ^2
\end{array}
\right\}
\longrightarrow\quad
\ima T^{J} = T^{J} \, \Sigma \, T^{J\,*}.
\label{uniT}
\end{equation}
In the last step we have reexpressed the unitarity relations in 
a matrix form, using
\be
T^{J}=\left(
\begin{array}{cc}
t^J_{\alpha\alpha}&t^J_{\alpha\beta}\\
t^J_{\alpha\beta}&t^J_{\beta\beta} \\
\end{array}
\right)
\quad ,\quad
\Sigma=\left(
\begin{array}{cc}
\sigma_\alpha&0\\
0 & \sigma_\beta\\
\end{array}
\right)\,,
\ee
which allows for a straightforward generalization to the
case of $n$ accessible states. A similar equation holds for multi-particle states,
but  we will comment about them only at the end of the section,
since both the expressions for the partial waves and the phase space factor 
are much more complicated.


Let us recall that, by writing 
$t^{J}_{\alpha\alpha}
=\vert t^{J}_{\alpha\alpha}\vert\exp(i\delta^{J}_{\alpha\alpha})$,
 eq.(\ref{uni1}) implies
the following bound 
\begin{equation}
\sigma_\alpha\,\vert t^{J}_{\alpha\alpha}\vert\le1 \qquad
\stackrel{s\rightarrow\infty}{\Longrightarrow}\qquad
\vert t^{J}_{\alpha\alpha}\vert \le1.
\label{naivebound}
\end{equation}
Note that in the high energy limit  
we recover the most familiar bound 
since $\sigma_\alpha\rightarrow1$ very rapidly when $s\rightarrow\infty$.
When a finite number of states are available, the $s\rightarrow\infty$
limit
provides also simple unitarity bounds. In particular,
the strongest one comes from the largest $T^{J}$ eigenvalue,
which again has to be smaller than one. This has been applied
to study the tree level unitarity bounds, at leading order in the
gauge couplings $g,g'$,
on the mass of the SM Higgs boson
in \cite{Lee:1977eg}. Following that example, in the neutral channel,
only three states are relevant, namely, $\alpha=W_{L}^+W_{L}^-$, 
$\beta=Z_{L}Z_{L}/\sqrt{2}$, $\gamma=HH/\sqrt{2}$ 
($HZ_{L}$ is decoupled from the others at tree level, 
and by itself yields
a smaller bound).
In the scalar, $J=0$ case,  
it is thus easy to calculate the 
$T^{J=0}$ eigenvalues in the $s\rightarrow\infty$ limit,
which should all be bounded by 1:
 \begin{equation}
   \label{eq:LQTmatrix}
T^{J=0}=\frac{G_F M_H^2}{4\pi\sqrt{2}}\left(
   \begin{array}{ccc}
1&1/\sqrt{8}&1/\sqrt{8}\\
1/\sqrt{8}&3/4&1/4\\
1/\sqrt{8}&1/4&3/4
   \end{array}
\right)\quad
\longrightarrow\quad
\frac{G_F M_H^2}{4\pi\sqrt{2}}(3/2,1/2,1/2)\le1,
 \end{equation}
As commented above, the largest one, 3/2, provides the stringent
unitarity bound: $M_H^2\le 8 \pi\sqrt{2}
/(3G_F)\simeq 2.7\pi\sqrt{2}/G_F$. With this
simple example it is already clear that by considering the coupled channel
unitarity relations, it is possible to find stronger
unitarity bounds than the naive one with a single channel,
$M_H^2\le 4\pi\sqrt{2}/G_F$.

However, the calculation of the determinant can be 
extremely  complicated either when the number of relevant coupled 
states is rather large or also at finite $s$, when the matrix 
elements are functions of $s$ instead of simple numbers. 
In particular, this is the case when we have a Higgs in the bulk
since the couplings to the higher KK scalar
field excitations
are not suppressed with $R$, but of the same order as the 
usual SM Higgs couplings. As a consequence, the tree level
scattering amplitudes between the
$HH$, $W_{L}^+W_{L}^-$ or $Z_{L}Z_{L}$ states, are of the same order 
as those considering their KK excitations. The latter states
are suppressed by phase space, since they are heavy, but 
not in the $s\rightarrow\infty$ limit where all of them
become relevant, and we end up with an intractable determinant.

In such a case there is an alternative method that one could 
use to obtain unitarity bounds that are somewhat weaker
than those obtained from the determinant but are much 
easier to calculate, and still stronger than those from the naive
single channel formalism. Let us go back to eq.(\ref{uniT})
and look at the equation for the diagonal element $t^J_{\alpha\alpha}$,
that we generalize to many accessible states $\alpha,\beta,\gamma,...$
as
\be
\ima t^J_{\alpha\alpha} =\sigma_\alpha \,\vert\, 
t^J_{\alpha\alpha} \vert ^2+ \sum_{\beta\neq\alpha}
\sigma_\beta \,\vert\, t^J_{\alpha\beta} \vert ^2,
\label{unitmany}
\ee
Of course, each state $\beta$ is not accessible below its threshold, 
and we define $\sigma_\beta=0$ if $s\leq s_{threshold\; of\;\beta}$.
By recalling again that 
$t^J_{\alpha\alpha}=\vert t^J_{\alpha\alpha}\vert\exp(i\delta^J_{\alpha\alpha})$
it is straightforward to arrive to the following bound:
\be
\hbox{Unit}_{\alpha\rightarrow\alpha}\equiv
\sigma_\alpha\vert t^J_{\alpha\alpha}\vert+
\frac{1}{\vert t^J_{\alpha\alpha}\vert}\sum_{\beta\neq\alpha}
\sigma_\beta \,\vert\, t^J_{\alpha\beta} \vert ^2 \le 1.
\label{newunitbound}
\ee
Let us remark, that all the terms in the sum {\it are positive},
and therefore this bound is always stronger than the naive one,
eq.(\ref{naivebound}). As a matter of fact, the sum in 
eq.(\ref{unitmany}) runs over all accessible states, but for
two-body states the partial waves are very easy to calculate
and that is why we consider them here. If one would like to 
include more states, the bounds would be even more stringent,
but the calculations extremely more cumbersome. Thus, we will 
restrict ourselves to the two-body contributions to this bound,
which, as we will see can already provide useful information.

As an example, the bound that we obtain
from the SM $T$ matrix in the $s\rightarrow\infty$ limit 
given in eq.(\ref{eq:LQTmatrix}), if we choose
$\alpha=W_{L}^+W_{L}^-$, is 
$M_H^2\le 16 \pi\sqrt{2}/(5G_F)\simeq 3.2\pi\sqrt{2}/G_F$,
much closer to the determinant bound than to the naive bound.
Of course, the real usefulness of this method comes into play when 
the determinant is hard to calculate, as we will show next
in the context of the SM extra dimensional extension.

\section{The 5D SM with the scalar sector in the bulk}
\label{section2}
 As a simple illustration of the use of the unitarity bounds
previously derived
let us consider a minimal 5D extension
of the SM  compactified on the segment $S^1/Z_2$, of length $\pi
R$, in which the $SU(2)_L$ and $U(1)_Y$ gauge fields and the
Higgs field $\Phi$   propagate in the bulk. The Lagrangian of the
gauge Higgs sector is given by (see \cite{Masip:1999mk})
\ba \int_{0}^{2 \pi R} dy\int
dx\,{\cal L}(x,y) &=&\int_{0}^{2 \pi R} dy\int dx\,\Big\{ - \frac
1 4 B_{MN}B^{MN} - \frac 1 4 F^a_{MN}F^{aMN}+{\cal L}_{GF}(x,y)
\nonumber\\
&+& (D_M \Phi)^\dagger (D^M \Phi)- V(\Phi) \Big\},
\label{kinlagrangian} 
\ea 
where $M=\mu,5$; $B_{MN}$, $F_{MN}^a$
are the $U(1)_Y$ and $SU(2)_L$ field strengths and
 $a$ is the $SU(2)$ index. The covariant derivative is defined as
$D_M=\partial_M - i g_5 A^a_M\tau^a/2 - i g_5'B_M/2$. We will
consider the following  Higgs potential
\begin{equation}
\label{scalarpotential} V(\Phi) = \mu^2 \, ( \Phi\da \Phi ) \, +
\, \lambda^{(5)} \, ( \Phi\da \Phi )^2.
\end{equation} 
The minimum of the potential corresponds
to the constant configuration $\Phi=(0,v/\sqrt{4\pi R})$,
 where $v^2\equiv -2 \pi R
\mu^2/\lambda^{(5)}=(\sqrt{2} G_F)^{-1}$. In this way, the Higgs field is expanded in
the standard form \be \Phi(x,y)=\left (
\begin{array}{c}
 \frac{i}{\sqrt{2}}(\omega^1-i\omega^2)\\
\dd { \frac{1}{\sqrt{2}}(\f {v}{\sqrt{2\pi R}}+ h-i \omega^3) }
\\
\end{array}
\right).
\ee

The gauge fixing Lagrangian is
\begin{eqnarray*}
{\cal L}_{GF}(x,y)=- \f {1}{2 \xi}
\left(\partial_\mu A^{a\,\mu }-\xi\left[\partial_5 A^a_5-
\f{g_5v }{2\sqrt{2\pi R}}\omega^a \right]\right)^2
- \f {1}{2 \xi}
\left(\partial_\mu B^{\mu }-\xi\left[\partial_5 B_5+ 
\f{g'_5v}{2\sqrt{2\pi R}}\omega^3\right]\right)^2
\end{eqnarray*}
where,
in order to avoid a gauge dependent mixing angle between
the physical $Z$ and the photon, we have chosen the same $\xi$ parameter
for the $A^{a\,\mu}$ and $B^\mu$ fields.
Let us now recall that the fields living in the bulk have a
Fourier expansion, which is:
\begin{eqnarray}
  X(x,y)=\f{1}{\sqrt{2\pi R}}X_{(0)}(x)+\f{1}{\sqrt{\pi R}}
\sum_{n=1}^{\infty}\cos\left(\f{n y}{R}\right)X_{(n)}(x),
\end{eqnarray}
for $X=A_\mu^a, B_\mu, \omega^a, h$,  whereas for $Y=A_5^a, B_5$
it is
\begin{equation}
    Y(x,y)=\f{1}{\sqrt{\pi R}}
\sum_{n=1}^{\infty}\sin\left(\f{n y}{R}\right)Y_{(n)}(x).
\end{equation}
After integration over the 5th  dimension,
the  Higgs fields have the following masses:
$m^2_{h(0)}=2v^2 \lambda$, $m^2_{h(n)}=m^2_{h(0)}+n^2/R^2$,
 where $\lambda=\lambda^{(5)}/(2\pi R)$.

Similarly to the SM case in four dimensions, we define the
following charged and neutral field combinations $W^{\pm}_M \, =
\lr{ A^1_M \, \mp \, i \, A^2_M } /\sqrt{2} $, $Z_M \,  = \lr{ g_5
\, A^3_M \, - \, g'_5 B_M }/\sqrt{g_5^2 + \gs}$, $A_M \,  = \lr{
g'_5 \, A^3_M \, + \, g_5 B_M }/\sqrt{g_5^2 + \gs}$. After
integrating out the compactified fifth dimension $y$, the mass
matrix of the gauge bosons and their KK excitations
is diagonal. Physically, this means that there is no mixing
between any KK mode of different KK level. One gets
 $m_{W(0)}=gv/2$, $m_{W(n)}=\sqrt{m_{W(0)}^2+n^2/R^2}$,
$m_{Z(0)}=\sqrt{g^2+g'^2}v/2$,
$m_{Z(n)}=\sqrt{m_{Z(0)}^2+n^2/R^2}$ with $g=g_5/\sqrt{2\pi R}$
and $g'=g'_5/\sqrt{2\pi R}$. The photon has zero mass and for its
associated KK states the masses are given by $m_{A(n)}=n/R$.

In terms of the KK modes, the gauge fixing conditions
become
\begin{eqnarray}
{\cal L}_{GF}(x)=   -\f{1}{\xi}\sum_{n=0}^{\infty}&& \left\{
\f{1}{2}\left[ \partial_\mu A^{\mu }_{(n)}-\xi
{\f{n}{R}}\,\,A_{(n)}^5\right]^2\right.+\left\vert
\partial_\mu  W^{+\,\mu }_{(n)}-
\xi m_{W(n)} G^+_{(n)}\right\vert^2\nonumber\\
&& \left. + \f{1}{2}\left[ \partial_\mu Z^{\mu }_{(n)}-\xi
m_{Z(n)} G^Z_{(n)} \right]^2\right\}.
\end{eqnarray}
where  we have defined
\begin{eqnarray}
&&G^\pm_{(0)}=-\omega^\pm_{(0)},\quad
G^\pm_{(n)}=
c^{_W}_{n} \, W^\pm_{5\,(n)}+ s^{_W}_n \,\omega^\pm_{(n)},
\; n\ge 1,\nonumber\\
&&G^Z_{(0)}=-\omega^3_{(0)},\quad
G_{(n)}^Z=
c^{_Z}_n \,\, Z_{5\,(n)}\,+ s^{_Z}_n \,\omega^3_{(n)}
, \,\; n\ge 1,
\label{Gs}
\end{eqnarray}
with $s^{_V}_n=-m_{V{(0)}}/m_{V{(n)}}$,
 $c^{_V}_n=(n/R)/ m_{V{(n)}}$ 
and $\omega^{\pm} = \frac{1}{\sqrt{2}}\lr{ \omega^1 \, \mp \, i
\, \omega^2 } $. Note that for brevity
we are using the notation $V=W^\pm, Z$.

 Once identified the pseudoscalar $G^V_{(n)}$
fields that couple diagonally with the derivatives of the gauge
boson mass eigenstates, the Equivalence Theorem \cite{ET,Lee:1977eg}
follows as usual also for the Kaluza-Klein gauge fields 
\cite{DeCurtis:2002nd}:
\begin{equation}
  T( V^{\mu }_{L\,(m)}, V^{\mu }_{L\,(n)},\ldots)
\simeq  C^{(m)} C^{(n)}... T( G^V_{(m)}, G^V_{(n)},\ldots)
+O(M_k/\sqrt{s}),
\end{equation}
$M_k$ being the biggest one of the
$m_{V(m)},m_{V(n)}...$ masses, and the $C^ {(i)}=1+O(g)$
account for renormalization corrections (see the last three references
in \cite{ET}). In this way, whenever we deal with scattering
amplitudes involving longitudinal
gauge bosons or their KK excitations, 
at energies much larger
than their masses,
we can simply calculate
using their corresponding pseudoscalar fields.
This is the reason why we have concentrated on 
partial waves of scalar fields.

 In general, for the
calculations of amplitudes we would also need the  orthogonal
combinations
\begin{eqnarray}
a^\pm_{(n)}=
-s^{_W}_{n} \, W^\pm_{5\,(n)}+ c^{_W}_n \,\omega^\pm_{(n)},\qquad
a_{(n)}^Z=
-s^{_Z}_n \,\, Z_{5\,(n)}\,+ c^{_Z}_n \,\omega^3_{(n)}
, \qquad n\ge 1.
\label{as}
\end{eqnarray}
with masses: $m^2_{a_{(n)}^V}=m_{V(0)}^2+n^2/R^2$.

In order to obtain the tree level unitarity bounds at the
lowest order in $g$ and $g'$,  we do not need the $\omega\omega
V$ couplings. Therefore, the only relevant interactions come from
the scalar potential, eq.(\ref{scalarpotential}).
After integrating out the fifth dimension,
for our calculation we need to recast this potential in
terms of the mass eigenstates, $G$ and
$a$, by means of
$\omega_{(0)}^V=-G^V_{(0)}$, $ \omega_{(n)}^V= c^{_V}_n
\,\, a^V_{(n)}\,- s^{_V}_n \,G^V_{(n)}$. Terms containing the
fields $G$ and $a$ come also from the $V_{5(n)}$ contributions in
the covariant derivative terms, but, as already stressed, they are
negligible at the lowest order in $g$ and $g'$. 
In \cite{DeCurtis:2002nd} we have calculated the
$W^{+}_{L\,(m)}W^{-}_{L\,(n)}\rightarrow
W^{+}_{L\,(p)}W^{-}_{L\,(q)}$ scattering amplitudes and shown that
the partial wave unitarity limit on the mass of the Higgs
receives only very tiny corrections from the pure KK gauge sector.
We want here to calculate the contribution to the tree level
unitarity limit from scattering amplitudes involving the scalars
$h_{(0)}$, $h_{(n)}$, $a^V_{(n)}$ and the longitudinal gauge bosons which are
related to the Goldstone bosons $G^V_{(0)}$ and  $G^V_{(n)}$ via
the ET.

\section{Unitarity bounds for the 5D SM with a Higgs in the bulk}

From the scalar potential  we can then calculate the
partial waves of the scattering of
the state $\alpha=W_{L\,(0)}^+W_{L\,(0)}^-$ into itself 
as well as into 
$\beta=h_{(0)} h_{(0)}/\sqrt{2}$,
$Z_{L\,(0)}Z_{L\,(0)}/\sqrt{2}$, $h_{(n)} h_{(n)}/\sqrt{2}$,
$a_{(n)}^3 a_{(n)}^3/\sqrt{2}$ and $a_{(n)}^+ a_{(n)}^-$, whose interactions are not
suppressed by any power of $R$ or $g$ and $g'$. Our
aim is to study the effect of the new KK states 
on the
tree level unitarity bounds on $M_H$, here called $m_{h(0)}$,
at leading order in $g$ and $g'$. As usual, we use the Equivalence Theorem \cite{ET}
to calculate the amplitudes replacing the longitudinal gauge
bosons by their associated Goldstone bosons. 
As it has been shown in \cite{DeCurtis:2002nd} the 
ET also holds for the KK excitations of longitudinal gauge bosons, 
which are associated to a combination of the fifth component 
of the gauge fields plus a part
from the Goldstone boson KK excitations, eq.(\ref{Gs}).
The latter is
 suppressed
by an $s^{_V}_n=-m_{V(0)}(R/n)  (1- m^2_{V(0)} R^2/(2
n^2)+\cdots)$ factor.
That is the reason why it is enough to 
consider the states previously mentioned and not the states
made of two longitudinal gauge boson
excitations or two transverse gauge boson components, 
since they are suppressed either by  $O(m_{V(0)}^2 R^2)$, $g$ or $g'$.
We give in the appendix the tree level scattering amplitudes
of $W^+_{L(0)}W^-_{L(0)}$ into the two body states $\beta$ mentioned
above, using the Equivalence Theorem.

First of all, let us emphasize that in the $s\rightarrow\infty$ limit,
the 5D SM violates 
tree level unitarity {\it for any value of the 
Higgs mass}. This is easily seen from eq.(\ref{newunitbound}),
since, for instance, in  the $s\rightarrow\infty$ limit
the amplitudes in the appendix are given by the quartic terms (trilinear terms
are suppressed by propagators)
\begin{eqnarray}
&&t^{J=0}_{W_ {L\,(0)}^+W_{L\,(0)}^-\rightarrow W_{L\,(0)}^+W^-_{L\,(0)}},\quad
t^{J=0}_{W_ {L\,(0)}^+W_{L\,(0)}^-\rightarrow a_{(n)}^+a^-_{(n)}}
\longrightarrow
-\frac{m_{h(0)}^2G_F\sqrt{2}}{8\pi},
\nonumber\\ \quad
&&t^{J=0}_{W_ {L\,(0)}^+W_{L\,(0)}^-\rightarrow Z_{L\,(0)}Z_{L\,(0)}},\quad
t^{J=0}_{W_ {L\,(0)}^+W_{L\,(0)}^-\rightarrow a_{(n)}^Za^Z_{(n)}}\longrightarrow
-\frac{m_{h(0)}^2G_F}{16\pi},
\nonumber\\
&&t^{J=0}_{W_{L\,(0)}^+W_{L\,(0)}^-\rightarrow h_{(0)} h_{(0)}},\qquad
t^{J=0}_{W_{L\,(0)}^+W_{L\,(0)}^-\rightarrow h_{(n)} h_{(n)}}
\longrightarrow
-\frac{m_{h(0)}^2G_F}{16\pi}.\quad
\label{sinftyamps}
\end{eqnarray}
Just by considering these states it is clear that the series on the
right hand side of  eq.(\ref{newunitbound}) $diverges$, for any value
of $m_{h(0)}$. All the other states that we have not considered {\it always 
add}, so that the series indeed diverges even more rapidly.

Of course, the $s\rightarrow\infty$ limit is not a problem if 
we consider the model as an effective theory, valid only up to 
some finite $s$. The tree level unitarity violation would then
show the point were perturbation theory breaks, 
or where the higher order loop
corrections become as large as the tree level ones and the theory
becomes strongly interacting.

Let us then illustrate the use of the unitarity bound derived in 
eq.(\ref{newunitbound}) at finite $s$, including several KK states
in the coupled channel formalism.  
We will give results for $\alpha=W^+_{L\,(0)}W^-_{L\,(0)}$
since we have checked that it
 provides the strongest bounds (as it also happened in the
$s\rightarrow\infty$
limit of the SM). 
As intermediate states we
use $\beta=h_{(0)} h_{(0)}/\sqrt{2},Z_{L\,(0)}Z_{L\,(0)}/\sqrt{2},h_{(n)} h_{(n)}/\sqrt{2},a_{(n)}^+ a_{(n)}^-/\sqrt{2}$ and
$a_{(n)}^Z a_{(n)}^Z/\sqrt{2}$ up to some given $n$.
In order to avoid problems
with the tree level propagators, which do not have widths
and can become infinite,  we consider  energies 
much larger than the masses involved in the calculation.
This choice also allows us to use the Equivalence Theorem as it 
was done in \cite{Lee:1977eg},
and substitute  each
$W_{L\,(0)}^\pm, Z_{L\,(0)}$ with its corresponding Goldstone bosons
$G^\pm_{(0)}$ and $G^Z_{(0)} $, which are, respectively,
$-\omega^\pm_{(0)}$ and $-\omega^3_{(0)}$.
As it was shown in \cite{DeCurtis:2002nd},
in the
small $R$ limit we have $c^{_V}_n=(1- m^2_{V(0)} R^2/(2
n^2)+\cdots)$. Hence, up to $O(m_{V(0)}^2R^2)$ corrections, we can also replace 
$a_{(n)}^\pm$ by $\omega_{(n)}^\pm$
and  $a_{(n)}^Z$ by  $\omega_{(n)}^3$.
All the partial waves
 needed for the calculation can be found in the appendix.

\begin{figure}[htbp]
\begin{center}

  \includegraphics[height=.21\textheight]{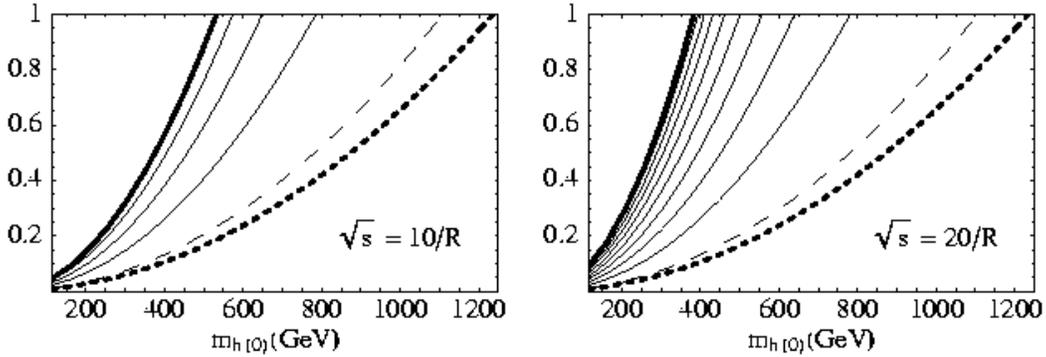}
  \caption{\small
The plots show for which
value of $m_{h(0)}$ the
tree level unitarity bound in eq.(\ref{newunitbound}) is violated.
The thick dotted line stands for
the SM ($n=0$) results 
only from the $W_ {L\,(0)}^+W_{L\,(0)}^-$ elastic scattering, 
eq.(\ref{naivebound}), whereas the dashed line 
includes also the $h_{(0)}h_{(0)}$,
$Z_ {L\,(0)}Z_{L\,(0)}$  coupled states in eq.(\ref{newunitbound}). 
The continuous lines correspond to 
considering in eq.(\ref{newunitbound}) the first, second, etc... 
KK excitations of the previous states.
The thick continuous line shows the complete calculation 
including all the
 kinematically allowed states, which,
for $\sqrt{s}=10/R$ and $20/R$, are
4 and 9 KK levels, respectively. 
Note that for {\it any} $m_{h(0)}$ there is 
a $\sqrt{s}$ above which tree level unitarity is violated.
The energies have been chosen below the
expected scale where the gauge coupling becomes non-perturbative, $30/R$
 \cite{Appelquist:2000nn}.}
\label{fig1:sm}
\end{center}
\end{figure}

In Figure 1 we show for what $m_{h(0)}$ value the unitarity bound in
eq.(\ref{newunitbound}) is violated
for a given energy and compactification radius. 
The curves represent the tree level approximation to the
left hand side of the inequality, so that
when they are larger than one show a violation of the unitarity bound
already by considering only two-particle states.
The thick dotted line corresponds to the analysis considering just the 
single $W_{L\,(0)}^+W_{L\,(0)}^-\rightarrow W_{L\,(0)}^+W_{L\,(0)}^-$ 
amplitude. 
The dashed line would be the bound including the other zero modes,
that is, the SM result for the left hand side of eq.(\ref{newunitbound}).
The continuous lines represent the change on 
eq.(\ref{newunitbound}) if one considers the coupled unitarity bound
including the first KK level, the second, etc... 
Of course, one should
consider all the KK states that can be produced up to
the energy under consideration, and that corresponds to
the thick continuous line.
We can notice that 
the unitarity bound can be reduced drastically by 
considering an additional extra dimension. 
Note that as we have previously shown,
for {\it any} $m_{h(0)}$ there is 
a $\sqrt{s}$ above which tree level unitarity is violated.

\begin{figure}[htbp]
\begin{center}
  \includegraphics[height=.40\textheight]{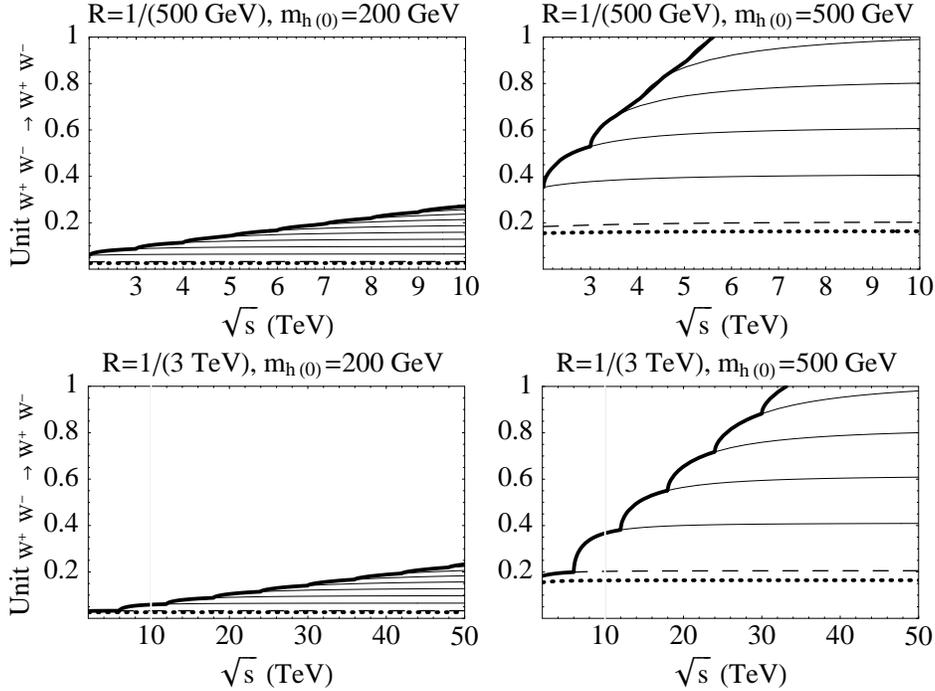}
  \caption{ \small The plots show the energies where the 
tree level calculation violates the coupled channel 
unitarity bound eq.(\ref{newunitbound}). The thick dotted
and dashed lines correspond to the SM bounds without KK excitations
just for the single channel, eq.(\ref{naivebound}), 
or coupled channel case, eq.(\ref{newunitbound}). Each continuous
line
represents the contribution of the new KK states that 
open as the energy increases. The total result
considering all the accessible KK level is the thick continuous line.}
\label{fig2}
\end{center}
\end{figure}

In Figure 2, we show  the energy at which 
tree level unitarity is violated for a given Higgs mass $m_{h(0)}$.
Roughly speaking, this means 
that beyond that point perturbation theory is no longer valid. 
However, let us remark that the saturation of the unitarity bounds
is also an indication that the model has become strongly interacting.
For practical purposes this can be considered the case when the tree
level calculation provides more than half of the unitarity bound.
Again the thick dotted and dashed lines correspond to considering
the SM fields in the single or coupled channel case, respectively, whereas
the continuous lines represent the contribution to the unitarity bound 
of each new level of KK accessible excitations at that energy.
The total result, considering all accessible KK lavels corresponds
to the thick continuous line.
Note that in the upper row of figures, we have chosen a 
compactification radius $R=1/(500\,$GeV$)$ and Higgs masses which
are within the presently $90\%$ allowed region in 5D
universal extra dimensional models
\cite{Barbieri:2000vh,Appelquist:2000nn}. 
The lower row, with $R=1/(3\,$TeV$)$
is the typical value for models with fermions localized on the brane
\cite{Masip:1999mk}.

In Figure 3, for different values of $R$, we show contour plots 
in the $\sqrt{s},m_{h(0)}$ plane, of 
Unit$_{W_ {L\,(0)}^+W_{L\,(0)}^-\rightarrow W_{L\,(0)}^+W_{L\,(0)}^-}$
in eq.{(\ref{newunitbound}). 
The white area represents the region
where the tree level calculation violates the unitarity bound
already considering two-particles states. 
Within the gray areas, the two-body amplitudes add to more than
half of the unitarity bound, suggesting that the theory becomes
strongly interacting.

Again, we can see in the $R=1/(500\,$GeV$)$ plots that 
the models can become strongly interacting within the reach of 
the LHC. This could be of relevance since the scale where
the gauge coupling 
becomes non-perturbative in 5D universal extra dimension models 
has been estimated around $30/R$ \cite{Appelquist:2000nn}. 
Within our approach it is not the running of the coupling, but
the proliferation of states and the fact that the coupling
is not suppressed when increasing the KK level, what drives the
tree level unitarity violation. If higher orders are to modify this behavior
they should be comparable to the tree level and the 
theory becomes strongly interacting.
As a matter of fact, we see in Fig.3 that the models
can become
strongly interacting much before the scale of 30/R,
depending on the Higgs mass. 
For instance, by looking at Figure 3 in the $R=1/(500 \mbox{GeV})$
case, the gauge coupling would become nonperturbative at 
approximately $30/R=15\,$TeV, but we see that 
for $m_{h(0)}=500\,$GeV  the theory violates tree level unitarity
already at 5 TeV and becomes strong below 3 TeV. 
We recall once again
that we are only considering two-particle states, 
but the many-particle states, since they always {\it add}
in eq.(\ref{newunitbound}), would even provide a stronger bound.

\begin{figure}[htbp]
  \begin{center}
    \includegraphics[height=.70\textheight]{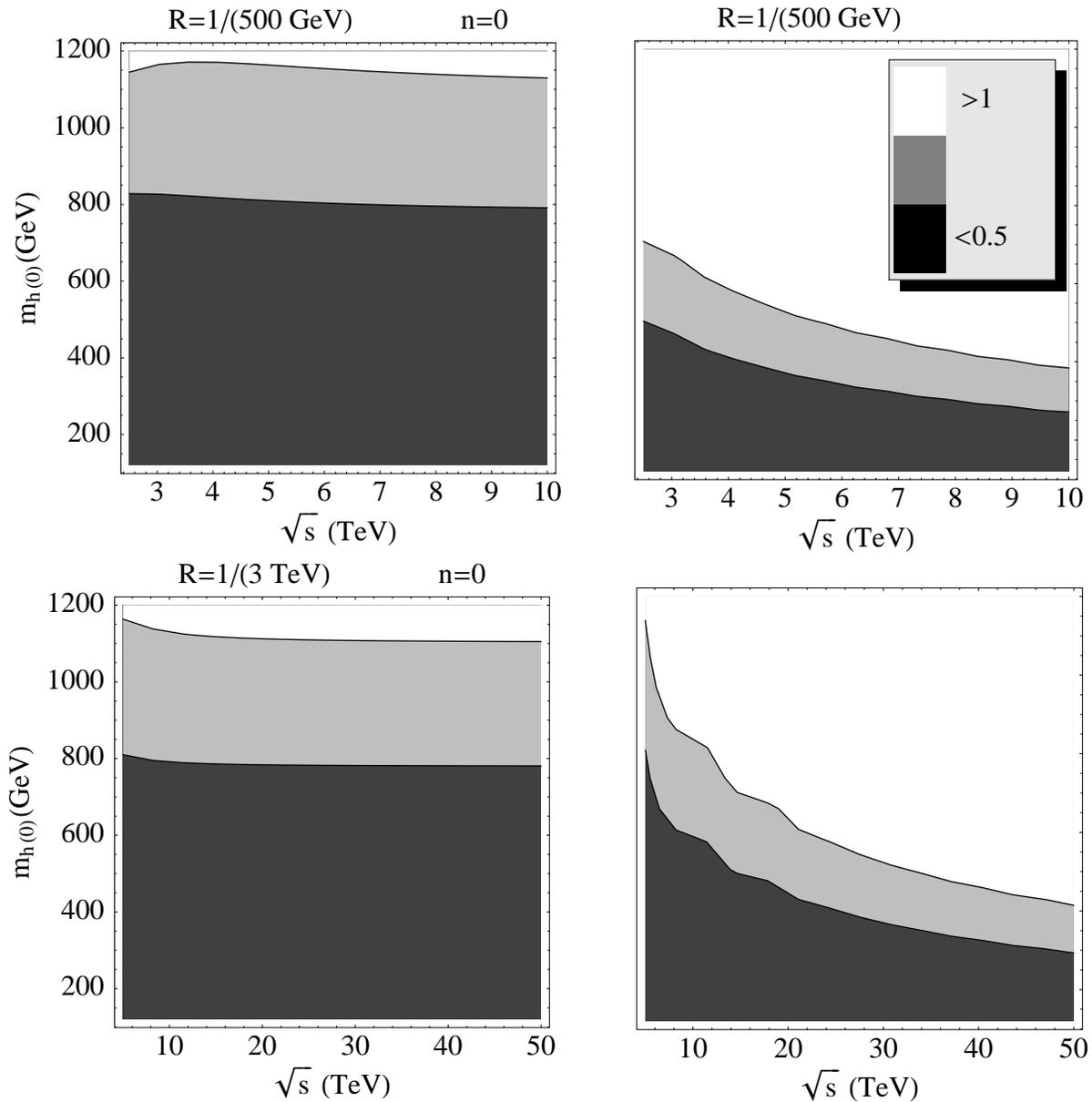}
  \caption{\small 
    The white areas represent the regions
of the $(\sqrt{s},m_{h(0)})$ plane
 where the tree
    level calculation violates the unitarity bound.  The gray areas
    represent the regions where 
Unit$_{W_ {L\,(0)}^+W_{L\,(0)}^-\rightarrow W_{L\,(0)}^+W_{L\,(0)}^-}$ in 
    eq.(\ref{newunitbound}) is larger than $0.5$, and suggest a
    strongly interacting regime.  We show,  the bounds
    obtained using only the SM fields ($n=0$), and those
    including the KK excitations. }
\end{center}
\label{fig3}
\end{figure}

Let us finally get a crude estimate of an energy for which
the tree level calculation violates unitarity for a given $m_{h(0)}$. 
It will not be as tight as the explicit calculation shown above
but in contrast will be very easy to implement.

First, if we
consider $s\gg m_{h(0)}^2$, we can 
approximate the modulus of the partial waves just by the constant
terms in eq.(\ref{sinftyamps}), which correspond to the quartic couplings.
That this is a fairly good approximation can be explicitly checked in the partial waves given
in the appendix,
since the trilinear terms
are suppressed by s-channel propagators, $1/(s-m_{h(0)}^2)$ 
or 
$m_{h(0)}^2/s$
factors in the logarithmic terms that appear from the 
angular integration of $t$ and $u$ channel propagators. 
The error in this approximation is $20\%$ for $\sqrt{s}\simeq5 m_{h(0)}$
and decreases very fast.
Remarkably, within this approximation, the partial waves
for KK modes are exactly a copy of the SM ones.
Each new KK level that opens up adds an additional copy.
Note that there is not any suppression in the amplitudes
when increasing the KK level.

Second, we have to count how many KK levels are 
effectively opened for a given energy. 
But for small  differences, all the new particles in a KK level
are characterized by a typical mass scale $m_{(n)}\simeq n/R>m_{h(0)}$.
As we have already commented, the two-particle phase space grows 
rather rapidly.
In particular, it can be checked that 
the phase space $\sigma_n=\sqrt{1-4m^2_{(n)}/s}$
of the two-particle state of the $n$-th
 KK level is of order one already at $\sqrt{s}> 3 m_{(n)}$.
Note that we are neglecting all the states that could have just opened at 
that energy but are below $\sqrt{s}> 3 m_{(n)}$ and would have contributed
 positively to the bound. 
Thus, at that energy, we have the following two-particle
states available:
the usual SM ones, plus $n$ KK copies, which as we have seen have 
the same amplitudes. 
For those energies, we can then approximate the sum of two-particle
states in  eq.(\ref{newunitbound}) as
\begin{equation}
  \label{eq:crude}
{\rm  Unit}_{W_ {L\,(0)}^+W_{L\,(0)}^-\rightarrow W_ {L\,(0)}^+W_{L\,(0)}^-}
\simeq \frac{5 m_h^2 G_F}{16 \pi \sqrt{2}}\sum_{k=0}^n \sigma_k
\simeq\frac{5 m_h^2 G_F}{16 \pi \sqrt{2}} (n+1)\leq1.
\end{equation}
Thus, we arrive to the following crude bound:
{\it given any value of the Higgs mass $m_{h(0)}$}, for energies
larger than
\begin{equation}
\sqrt{s}\simeq
\frac{3}{R}
\left[\frac{16\pi\sqrt{2} }{5\, G_F\, m^2_{h(0)}}-1\right]\quad (s\gg m^2_{h(0)}),
\label{crudeestimate}
\end{equation}
tree level unitarity is violated. This bound is only applicable
for $\sqrt{s}> m_{h(0)}$.y
We have checked that this formula gives a 
reasonably accurate bound within $10\%$ of the complete calculation. 
Indeed, we see by comparing Fig.4.a with 3.b
that the 5D results obtained with the estimate, eq.(\ref{crudeestimate}),
are in good agreement with the complete tree level calculation 
eq.(\ref{newunitbound}).

\section{Additional extra dimensions}

In the previous section we have
obtained very strong
tree level unitarity bounds 
when we added to the SM an additional tower of KK states coming
from a compactified extra dimension.
These contributions are always present when
new states become available, and 
if the couplings to these new states are not suppressed
when increasing the KK level, they can be comparable to those
of the SM fields.
We have seen that for sufficiently high energies and
just one extra dimension, the number of two-particle states
grows linearly with $\sqrt{s}$. However, we will see next
that when considering more than one extra dimension, the
number of states grows much faster, and the unitarity
bounds become extraordinarily much tighter.
This result is of relevance in the context of universal extra
dimensions \cite{Appelquist:2000nn}, where the problem of proton
instability has been solved precisely 
in six dimensions \cite{Appelquist:2001mj}.
 
This is particularly simple to see if we 
just add another dimension with the same compactification
radius to the previous model. 
In such a case, instead of a generic $\alpha_{(n)}$ tower of states
of two particles with mass $m_{(n)}\simeq n/R$,
we have $\alpha_{(n_1,n_2)}$ states 
of two particles with mass 
$m_{(n_1,n_2)}\simeq\sqrt{n_1^2+n_2^2}/R$.
Note that we are now writing explicitly the KK level,
because in the models of interest there is no mixing
among the modes with different KK numbers.

As before we are interested in two-body states that
can  be produced 
at tree level from
a state with two zero-level particles. 
In Table 1, we find the states
made of two particles with the same KK numbers, available
as the level number
$n\ge n_1,n_2$ increases.

\begin{table}[htbp]
\centering
\begin{tabular}[h]{|l|l|l|}
\hline
KK level&5D&6D\\\hline
0th& $\alpha_{(0)}$ & $\alpha_{(0,0)}$\\\hline
1st& $\alpha_{(1)}$ & $\alpha_{(1,0)}$, 
$\alpha_{(0,1)}$, $\alpha_{(1,1)}$\\\hline
2nd& $\alpha_{(2)}$ & $\alpha_{(2,0)}$, 
$\alpha_{(0,2)}$, $\alpha_{(2,1)}$, 
$\alpha_{(1,2)}$, $\alpha_{(2,2)}$\\\hline
3rd& $\alpha_{(3)}$ & $\alpha_{(3,0)}$, 
$\alpha_{(0,3)}$, $\alpha_{(3,1)}$, 
$\alpha_{(1,3)}$, $\alpha_{(2,3)}$, $\alpha_{(3,2)}$, $\alpha_{(3,3)}$\\
\hline
4th &$\alpha_{(4)}$& 9 states\\\hline
... &...& ...\\\hline
nth&$\alpha_{(n)}$&2n+1 states\\\hline
\end{tabular}
\caption{States made of two particles with the same KK numbers,
available as each KK level opens
for  one or two $S^1/Z_2$ extra dimensions in the model of section 3.
 }
\label{tab:levels}
\end{table}
We see that for the 5D case
we had $n+1$ states at the $n$th level
whereas  
for the 6D case we have $(n+1)^2$. 
Note that for $n\ge3$, when the $(n,n)$ state is opened there can also be
additional opened states from the $n+1$ level, for instance the
 $(n+1,0)$, which is lighter than $(n,n)$. But considering
only those up to $n\ge n_1,n_2$ we can easily count the number of states
and thus reobtain  the crude estimate in eq.(\ref{crudeestimate})
changing the number of states available
\begin{equation}
  \label{eq:crude6d}
  {\rm Unit}_{W_ {L\,(0)}^+W_{L\,(0)}^-\rightarrow W_ {L\,(0)}^+W_{L\,(0)}^-}\simeq
\frac{5 m_h^2 G_F}{16 \pi \sqrt{2}} (n+1)^2\leq1.
\end{equation}
As usual, we have restricted ourselves to two-body states because they
are very easy to calculate,
but all the other accessible states
that we have not counted always {\it add} to the
equation above and hence would have given an even stronger bound
on the Higgs mass.
Next,
we note that the phase space of the heaviest 
two-particle $nth$-level state, which is 
made of two $\alpha_{(n,n)}$  fields of mass $m_{(n,n)}\simeq\sqrt{2}n/R$,
is of order one when $\sqrt{s}\simeq 3m_{(n,n)}/R$ or larger. Thus, we now
expect a violation for energies above or around
\begin{equation}
  \sqrt{s}\simeq
\frac{3\sqrt{2}}{R}\left[\sqrt{\frac{16\pi\sqrt{2} }{5\, G_F\, m^2_{h(0)}}}-1\right].
\label{crudeestimate6d}
\end{equation}
Hence,
in this case we can consider eq.(\ref{crudeestimate6d}) more conservative
than the analogous in 5D, eq.(\ref{crudeestimate}).
If we had also considered
opened states beyond the $n$th level, like $(n+1,0)$,
 the unitarity bound above would even be tighter 
(but the counting of states would
be more complicated). 

The generalization to $d$ additional dimensions is straightforward
by changing $(n+1)^2$ by $(n+1)^d$ in eq.(\ref{eq:crude6d}).
Of course, the model could be more complicated and have different
compactification radius for each dimension, in such a case, if for
a given energy the KK states associated to the first extra dimensions
are accessible up to the $n_1$ level and those of the next up to $n_2$ , etc...
we have to substitute $(n+1)^d$ by $(n_1+1)(n_2+1)...(n_d+1)$.

\begin{figure}[htbp]
  \begin{center}
    \includegraphics[height=.23\textheight]{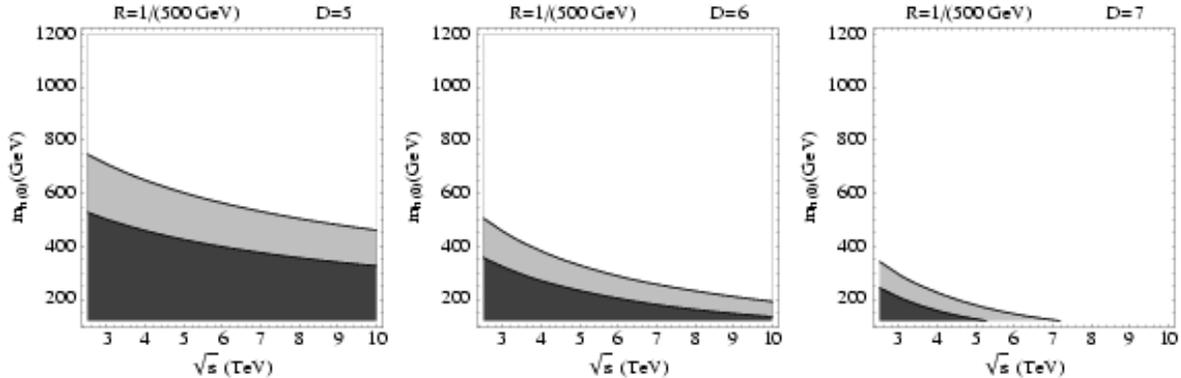}
  \caption{\small 
    The white areas represent the regions 
of the $(\sqrt{s},m_{h(0)})$ plane
where the tree
    level estimate, eqs.(\ref{crudeestimate}),(\ref{crudeestimate6d}),
violate the unitarity bound for 5, 6 or 7 dimensions.  The gray areas
    represent the regions where Unit$_{\alpha\rightarrow\alpha}$ in 
    eq.(\ref{newunitbound}) is larger than $0.5$, and suggest a
    strongly interacting regime. By comparing Fig.4.a with 3.b
it can be noticed that the 5D results obtained with the estimate
are in good agreement with the complete tree level calculation.}
\end{center}
\label{fig4:diagrams}
\end{figure}

In Fig.4 we show the results of using the crude estimate 
for the tree level unitarity violation, 
eqs.(\ref{crudeestimate}) and (\ref{crudeestimate6d}).
For illustration we have chosen again a value of $1/R$ 
which is presently allowed in the 5D and 6D universal extra dimension
context. For the 7D results we have simply replaced $(n+1)^2$
by $(n+1)^3$ in eq.(\ref{crudeestimate6d}) as explained above.
It can be noticed that by increasing the number of dimensions
where the scalar sector lives, the violation of tree level unitarity can be
dramatic within the LHC reach, which could be of considerable
phenomenological relevance.

\section{Conclusions}

In this work we have presented a very simple method to obtain unitarity bounds
at finite energy when a large number of states are available.
These kind of bounds are very relevant in order to determine
the energy at which perturbation theory breaks down for a given
set of parameters of the theory. These bounds also suggest the condition
under which the theory becomes strongly interacting.
The approach relies on the well known
formalism for coupled channel partial 
wave unitarity. Thus, it can be applied generically 
 and avoids the calculation of large determinants
of complicated functions. 

Our initial motivation to look for this kind of method are the extra 
dimensional
extensions of the SM, which introduce an infinite tower of Kaluza Klein states
that could saturate the unitarity bounds much before than in the familiar 
four-dimensional
SM. 
Although the bounds obtained from this formalism can be applied
to particles with any spin, we have illustrated this effect with 
a SM whose scalar sector is located in the bulk.
Incidentally this one as well as other similar models where the 
self-couplings of the scalar potential are not 
suppressed for higher Kaluza Klein states
have received some recent phenomenological interest.

We have indeed shown that within these models, shortly after a new KK level
can be produced, their contribution to the unitarity bound is of the same order
of that of the SM, thus changing the usual bounds dramatically as soon as 
several KK level are opened. In certain cases, this would imply that the 
model becomes strongly interacting much before the usual expectations,
thus casting doubts about simple perturbative calculations or estimates.
Let us finally remark the possible phenomenological interest of these bounds,
since within some valid regions of parameter space, 
they  could well suggest a strongly interacting regime within the 
reach of the next generation of colliders.

\appendix

\section{Appendix}

To study the tree level unitarity
bounds we are interested  in the
scattering of longitudinal gauge bosons
$W^+_{L(0)}W^-_{L(0)}$ at leading order in $g$ and $g'$.
In the SM at tree level this state is also coupled
to $Z_{L(0)}Z_{L(0)}$ and $h_{(0)}h_{(0)}$.
Using the ET, at high energies, $\sqrt{s}>m_{V(0)}$ the longitudinal
gauge boson amplitudes can be replaced by their
corresponding Goldstone boson amplitudes.
Using the 5D potential in eq.(\ref{scalarpotential}) 
and integrating out the fifth dimension as explained in the text,
we find
in the Landau gauge $\xi\rightarrow1$:
\begin{eqnarray}
  \label{eq:wpwmwpwm}
  t^{J=0}_{G^+_{(0)}G^-_{(0)}\rightarrow G^+_{(0)}G^-_{(0)}}
=\frac{- G_F m^2_{h(0)}}{8 \pi \sqrt{2}}\left[
2+\frac{ m^2_{h(0)}}{s- m^2_{h(0)}}-\frac{m^2_{h(0)}}{s}\log\Big(
1+\frac{s}{m^2_{h(0)}}\Big)\right],
\end{eqnarray}
\begin{eqnarray}
  \label{eq:wpwm33}
  t^{J=0}_{G^+_{(0)}G^-_{(0)}\rightarrow G^Z_{(0)}G^Z_{(0)}}
=\frac{- G_F\, m^2_{h(0)}}{16 \pi }\left[
1+\frac{ m^2_{h(0)}}{s- m^2_{h(0)}}\right],
\end{eqnarray}
\begin{eqnarray}
  \label{eq:wpwm00}
  t^{J=0}_{G^+_{(0)}G^-_{(0)}\rightarrow h_{(0)}h_{(0)}}
=\frac{- G_F\, m^2_{h(0)}}{16 \pi}\left[
1+\frac{ 3m^2_{h(0)}}{s- m^2_{h(0)}}+
\frac{4m^2_{h(0)}}{s\,\sigma_{h(0)}}\log\Big(
\frac{s-2m^2_{h(0)}-s\,\sigma_{h(0)}}{2m^2_{h(0)}}\Big)\right],
\end{eqnarray}
where $\sigma_\Phi=\sqrt{1-4m^2_\Phi/s}$
and we have already included the $1/\sqrt{2}$ factor for identical particles.
Incidentally, they are the same as in the SM \cite{Lee:1977eg}.

In addition, in 5D the longitudinal gauge bosons also couple
to their KK excitations
as well as to $h_{(n)}h_{(n)}$,
$a^+_{(n)}a^-_{(n)}$,
$a^Z_{(n)}a^Z_{(n)}$. In \cite{DeCurtis:2002nd} we have
already shown that the corrections to the SM results
from the longitudinal KK gauge excitations
are suppressed as $O(m_W^4/R^4)$ and can be also neglected on
a first approximation. However, we need the following
partial waves,
\begin{eqnarray*}
  \label{eq:wpwmapam}
  t^{J=0}_{G^+_{(0)}G^-_{(0)}\rightarrow a^+_{(n)}a^-_{(n)}}
=\frac{- G_F m^2_{h(0)}}{8 \pi \sqrt{2}}\left[
2+\frac{ m^2_{h(0)}}{s- m^2_{h(0)}}-\frac{m^2_{h(0)}}
{s\sigma_{a^\pm(n)}}\log\left(
\frac{2m^2_{a^\pm(n)} -2m^2_{h(n)}-s(1+\sigma_{a^\pm(n)}) }
{ 2m^2_{a^\pm(n)} -2m^2_{h(n)}-s(1-\sigma_{a^\pm(n)}) }\right)\right],
\end{eqnarray*}
\begin{eqnarray*}
  \label{eq:wpwpa0a0}
  t^{J=0}_{G^+_{(0)}G^-_{(0)}\rightarrow a^Z_{(n)}a^Z_{(n)}}
=\frac{- G_F\, m^2_{h(0)}}{16 \pi }\left[
1+\frac{ m^2_{h(0)}}{s- m^2_{h(0)}}\right],
\end{eqnarray*}
\begin{eqnarray*}
  \label{eq:wpwmnn}
  t^{J=0}_{G^+_{(0)}G^-_{(0)}\rightarrow h_{(n)}h_{(n)}}
=\frac{- G_F\, m^2_{h(0)}}{16 \pi}\left[
1+\frac{ 3m^2_{h(0)}}{s- m^2_{h(0)}}-
\frac{2m^2_{h(0)}}{s\,\sigma_{h(n)}}
\log\left(
\frac{2m^2_{h(n)} -2m^2_{a^\pm(n)}-s(1+\sigma_{h(n)}) }
{ 2m^2_{h(n)} -2m^2_{a^\pm(n)}-s(1-\sigma_{h(n)}) }\right)\right].
\end{eqnarray*}
Note that all these amplitudes for the $a^V_{(n)}$
are obtained replacing $ \omega_{(n)}^V\rightarrow c^{_V}_n
\,\, a^V_{(n)}$ in the Higgs potential
after the integration of the 5th dimension.
When deriving these amplitudes we have considered
the leading order in $c^{_V}_n=(1- m^2_{V(0)} R^2/(2
n^2)+\cdots)$ , thus neglecting $O(m_{V(0)}^2R^2)$.
In the worst case considered in our calculations 
$m_{V(0)}^2R^2\simeq (90/500)^2$, less than a 4\%.

\section*{Acknowledgments}

J.R.P. thanks
support from the Spanish CICYT projects
PB98-0782 and BFM2000 1326, as well as a
Marie Curie fellowship MCFI-2001-01155 .


\footnotesize
\begin{thebibliography}{99}




\bibitem{Antoniadis:1990ew}
I.~Antoniadis,
Phys.\ Lett.\ B {\bf 246} (1990) 377;
A.~Pomarol and M.~Quir\'os,
Phys.\ Lett.\ B {\bf 438} (1998) 255; 
I.~Antoniadis, S.~Dimopoulos, A.~Pomarol and M.~Quir\'os,
Nucl.\ Phys.\ B {\bf 544} (1999) 503; 
A.~Delgado, A.~Pomarol and M.~Quir\'os,
Phys.\ Rev.\ D {\bf 60} (1999) 095008. 



\bibitem{Masip:1999mk}
I.~Antoniadis and K.~Benakli,
Phys.\ Lett.\ B {\bf 326} (1994) 69;
P.~Nath and M.~Yamaguchi,
Phys.\ Rev.\ D {\bf 60} (1999) 116004;
W.~J.~Marciano,
Phys.\ Rev.\ D {\bf 60} (1999) 093006;
M.~Masip and A.~Pomarol,
Phys.\ Rev.\ D {\bf 60} (1999) 096005; 
T.~G.~Rizzo and J.~D.~Wells,
Phys.\ Rev.\ D {\bf 61} (2000) 016007; 
A.~Strumia,
Phys.\ Lett.\ B {\bf 466} (1999) 107; 
R.~Casalbuoni, S.~De Curtis, D.~Dominici and R.~Gatto,
Phys.\ Lett.\ B {\bf 462} (1999) 48; 
C.~D.~Carone,
Phys.\ Rev.\ D {\bf 61} (2000) 015008; 
A.~Delgado, A.~Pomarol and M.~Quir\'os,
JHEP {\bf 0001} (2000) 030.
A.~Muck, A.~Pilaftsis and R.~Ruckl,
Phys.\ Rev.\ D {\bf 65} (2002) 085037.


\bibitem{Barbieri:2000vh}
R.~Barbieri, L.~J.~Hall and Y.~Nomura,
Phys.\ Rev.\ D {\bf 63} (2001) 105007; 

\bibitem{Appelquist:2000nn}
T.~Appelquist, H.~C.~Cheng and B.~A.~Dobrescu,
Phys.\ Rev.\ D {\bf 64} (2001) 035002.
T.~Appelquist and H.~U.~Yee,
Phys.\ Rev.\ D {\bf 67} (2003) 055002.


\bibitem{SekharChivukula:2001hz}
R.~S. Chivukula, D.~A.~Dicus and H.~J.~He,
Phys.\ Lett.\ B {\bf 525} (2002) 175; 
R.~S.~Chivukula and H.~J.~He,
Phys.\ Lett.\ B {\bf 532} (2002) 121.

\bibitem{ET} 
J. M. Cornwall, D. N. Levin and G. Tiktopoulos,
\PR{D10}   (1974) 1145;
C. E. Vayonakis,  Lett. Nuovo Cim. {\bf 17}  (1976) 383;
M. S. Chanowitz  and M. K. Gaillard \NP{B261}
  (1985) 379; G. K. Gounaris, R. Kogerler and H. Neufeld,
\PR{D34}   (1986) 3257; Y. P. Yao and C. P. Yuan, \PR{D38}  (1988)  2237;
J. Bagger and C. Schmidt, \PR {D41} (1990)  264;
H.J. He, Y. P. Kuang, and X. Li \PRL{69} (1992) 2619.


\bibitem{Lee:1977eg}
B.~W.~Lee, C.~Quigg and H.~B.~Thacker,
Phys.\ Rev.\ D {\bf 16} (1977) 1519.


\bibitem{DeCurtis:2002nd}
S.~De Curtis, D.~Dominici and J.~R.~Pelaez,
Phys. Lett. B {\bf 554} (2003) 164.

\bibitem{scattering}
R.G. Newton,{\it''Scattering Theory of waves and Particles''},
2nd ed. Texts and Monographs in Physics. Springer Verlag NY (1982).
A.~Dobado, A.~Gomez-Nicola, A.~Maroto and J.~R.~Pelaez,
{\it``Effective Lagrangians For The Standard Model''},
 Texts and Monographs in Physics. Springer Verlag NY (1997).

\bibitem{Appelquist:2001mj}
T.~Appelquist, B.~A.~Dobrescu, E.~Ponton and H.~U.~Yee,
Phys.\ Rev.\ Lett.\  {\bf 87} (2001) 181802
\end{thebibliography}
\end{document}